\documentclass[aps,preprint,amsmath,amssymb,showpacs,amsfonts]{revtex4}
\usepackage{epsfig}
\usepackage{graphicx}
\usepackage{dcolumn}
\usepackage{bm}

\newcommand{\beq}{\begin{equation}}
\newcommand{\eeq}{\end{equation}}
\newcommand{\bea}{\begin{eqnarray}}
\newcommand{\eea}{\end{eqnarray}}

\begin{document}

\title{Relativistic CFT Hydrodynamics from the Membrane Paradigm}

\author{Christopher Eling$^1$}
\author{Yaron Oz$^2$}
\affiliation{$^1$ Racah Institute of Physics,
  Hebrew University of Jerusalem,
  Jerusalem 91904, Israel}
  \affiliation{$^2$ Raymond and Beverly Sackler School of
Physics and Astronomy, Tel-Aviv University, Tel-Aviv 69978, Israel}

\date{\today}

\begin{abstract}
We use the membrane paradigm to analyze the
horizon dynamics of a uniformly boosted black brane in a $(d+2)$-dimensional
asymptotically Anti-de-Sitter space-time and a Rindler acceleration horizon in $(d+2)$-dimensional  Minkowski space-time. We
show that in these cases the horizon dynamics is governed by the relativistic CFT
hydrodynamics equations. The fluid velocity and temperature correspond to the normal
to the horizon and to the surface gravity,
respectively. The second law of thermodynamics for the fluid is
mapped into the area increase theorem of General Relativity. The
analysis is applicable, in general, to perturbations around a stationary horizon, when the scale of variations of the macroscopic
fields is much larger than the inverse of the temperature.  We show that  the
non-relativistic limit of our analysis yields the incompressible Navier-Stokes equations.

\end{abstract}
\pacs{04.70.-s, 47.10.ad, 11.25.Tq }
\maketitle

\section{Introduction}

The hydrodynamics of relativistic conformal field theories (CFTs)
has attracted much attention recently, largely in view of the
AdS/CFT correspondence between gravitational theories on
asymptotically Anti-de-Sitter (AdS) spaces and CFTs
\cite{Maldacena:1997re} (for a review see \cite{Aharony:1999ti}).
Hydrodynamics is an effective description of the long distance field
theory dynamics and applies under the condition that the correlation
length of the fluid $l_{cor}$ is much smaller than the
characteristic scale $L$ of variations of the macroscopic fields.
The AdS/CFT correspondence suggests that the long wavelength
dynamics of gravity provides a dual description of the CFT
hydrodynamics.

It has been shown in \cite{Bhattacharyya:2008jc} that the
$(d+1)$-dimensional CFT hydrodynamics equations are the same as the
equations describing the evolution of large scale perturbations of
the $(d+2)$-dimensional black brane. The derivation of this result
parallels the conventional derivation of hydrodynamics equations
from the Boltzmann equation \cite{Fouxon:2008ik}, where the
``thermal equilibrium" solution is the boosted black brane. Thus,
the equations of gravity play for the hydrodynamic equations of a
strongly coupled CFTs the same role as the Boltzmann equation plays
at a weak coupling.

The limit of non-relativistic macroscopic motions in CFT
hydrodynamics leads to the non-relativistic incompressible
Navier-Stokes (NS) equations \cite{Fouxon:2008tb,
Bhattacharyya:2008kq}. Since we can obtain the NS equations in the
non-relativistic limit of CFT hydrodynamics, the AdS/CFT
correspondence implies that these equations have a dual gravitational
description, which can found by taking the non-relativistic limit of the
geometry dual to the relativistic CFT hydrodynamics
\cite{Bhattacharyya:2008kq}.

In this duality picture the dynamics of the fluid entropy of the field theory
at the asymptotic boundary can be expressed directly in terms of the black brane horizon geometry \cite{Bhattacharyya:2008xc}.
This behavior of the horizon is reminiscent of the membrane paradigm
in classical General Relativity, where the dynamics of the
black hole event horizon is analogous to that of a fictitious fluid
\cite{Damour,Price:1986yy,membrane}. Indeed, the membrane paradigm
has already been an important tool for calculating transport coefficients of the boundary gauge theory \cite{Kovtun:2003wp, Saremi:2007dn, Fujita:2007fg, Starinets:2008fb, Iqbal:2008by}.

In \cite{Eling:2009pb} we used the membrane paradigm formalism and
an expansion in powers of the Knudsen number $l_{cor}/L$ to show that the dynamics of a membrane defined by the event horizon of a black brane in asymptotically AdS space-time is described by the incompressible
Navier-Stokes equations of non-relativistic fluids. Moreover, the
analysis performed in \cite{Eling:2009pb} holds for a non-singular
null hypersurface, provided a large scale hydrodynamic limit exists.

The purpose of this paper is to generalize the analysis of
\cite{Eling:2009pb} to the relativistic CFT hydrodynamics. Our starting point is an equilibrium $(d+2)$-dimensional solution containing a timelike Killing vector field and a stationary $(d+1)$-dimensional causal horizon. This solution is associated with a thermal state at uniform temperature. When a hydrodynamic limit exists, we can expand the solution in the neighborhood of the causal horizon in powers of $l_{cor}/L$. We consider two specific examples in this paper: a black brane in asymptotically AdS and a Rindler acceleration horizon in Minkowski space-time. Assuming there is no singularity at the horizon, we show that at lowest orders in $l_{cor}/L$ the set of Einstein equations projected into the horizon surface is equivalent to the $(d+1)$-dimensional relativistic CFT Navier-Stokes equations. Our results imply that the analogy between horizon dynamics and hydrodynamics in the membrane paradigm is in fact an identity in certain cases. Since the non-relativistic incompressible NS equations arise in the slow motion limit of CFT hydrodynamics
\cite{Fouxon:2008tb,Bhattacharyya:2008kq}, we will also obtain
results of \cite{Eling:2009pb} in this limit.

The paper is organized as follows. In Section 2 we will briefly
review some of the basics of CFT hydrodynamics that we will need,
including the expansion in the Knudsen number and the form of the
stress-energy tensor in the ideal and dissipative cases. In Section
3 we will outline the membrane paradigm formalism and consider the
geometry and dynamics of null hypersurfaces and of the stretched
horizon. We also clarify the relationship between the membrane
dynamics in general and real hydrodynamics. In Section 4 we will
apply the membrane paradigm to a uniformly boosted black brane in a
($d+2$)-dimensional asymptotically AdS space-time and show that the
horizon dynamics is governed by the CFT hydrodynamics equations. In
Section 5 we will show that our analysis is applicable to cases
other than the black branes in AdS by considering the example of
Rindler space associated with accelerated observers in $d+2$
Minkowski space-time. We will show that the Rindler horizon dynamics
is also governed by the CFT hydrodynamics equations. In Section 6
will consider the non-relativistic limit of our analysis and
re-derive the non-relativistic hydrodynamics results of
\cite{Eling:2009pb}. Along the way we will clarify the relation
between the surface gravity and the fluid pressure. We conclude in
Section 7 with a discussion of open problems. In the following,
unless explicitly stated, we will use the convention $G=c=\hbar=k_B=1$.

\section{Relativistic CFT hydrodynamics}

Conformal hydrodynamics in $(d+1)$-dimensional space-time ($d \geq
2$) is described by $d+1$ fields:  temperature $T(x)$ and the
$(d+1)$-velocity vector field $u^{\mu}(x), \mu=0,...,d$, satisfying
$u_{\mu}u^{\mu}=-1$. The stress-energy tensor of the CFT obeys
\begin{equation}
\partial_{\nu}T^{\mu\nu}=0,~~~~~T^{\mu}_{\mu}=0 \ ,
\label{cfteq}
\end{equation}
and the equations of relativistic hydrodynamics are determined by
the constitutive relation expressing $T^{\mu\nu}$ in terms of the
temperature and the four-velocity field. The constitutive relation
has the form of a series in the small parameter (Knudsen number)
\begin{equation}
Kn\equiv l_{cor}/L \ll 1 \ ,
\end{equation}
where $l_{cor}$ is the correlation length of the fluid and $L$ is
the scale of variations of the macroscopic fields. Since the only
dimensionfull parameter is the characteristic temperature of the
fluid $T$, one has by dimensional analysis that $l_{cor} \sim
\frac{1}{T}$. The constitutive relation reads
\begin{eqnarray}&&
T^{\mu\nu}(x)=\sum_{l=0}^{\infty}T^{\mu\nu}_l(x),\ \ T^{\mu\nu}_l\sim (Kn)^l, \label{series}
\end{eqnarray}
where $T^{\mu\nu}_l(x)$ is determined by the local values of
$u^{\mu}$ and $T$ and their derivatives of a finite order. Keeping
only the first term in the series gives ideal hydrodynamics, while
dissipative hydrodynamics arises when one keeps the first two terms
in the series.

The ideal hydrodynamics approximation for $T^{\mu\nu}$ does not
contain the spatial derivatives of the fields. The $l=0$ term in
(\ref{series}) gives the stress-energy tensor that reads (up to a
multiplicative constant)
\begin{eqnarray}&&
T_{\mu\nu}= T^{d+1}\left[\eta_{\mu\nu}+(d+1)u_{\mu}u_{\nu}\right] \ ,
\label{ideal00}
\end{eqnarray}
where $\eta_{\mu\nu} = diag[-,+,+,..,+]$.

The dissipative hydrodynamics is obtained by keeping the $l=1$ term
in the series in Eq.~(\ref{series}). In the Landau frame
\cite{Landau,Bhattacharyya:2008jc,Baier:2007ix}, that fixes the
ambiguity in the form of the stress-energy tensor under a field
redefinition of the temperature and velocity, the stress-energy
tensor reads (up to a multiplicative constant)
\begin{eqnarray}&&
T_{\mu\nu}=T^{d+1}\left[\eta_{\mu\nu}+(d+1)u_{\mu}u_{\nu}\right]-2\eta
\sigma_{\mu\nu}, \label{visc00}
\end{eqnarray}
where the shear tensor $\sigma_{\mu\nu}$ obeys
$\sigma_{\mu\nu}u^{\nu}=0$ and is given by
\begin{eqnarray}&&
\sigma_{\mu\nu}=\frac{1}{2}\left(\partial_{\mu} u_{\nu}+
\partial_{\nu} u_{\mu}+u_{\nu}u^{\rho}\partial_{\rho} u_{\mu}+
u_{\mu}u^{\rho}\partial_{\rho} u_{\nu}\right)
-\frac{1}{d}\partial_{\alpha}u^{\alpha}
\left[\eta_{\mu\nu}+u_{\mu}u_{\nu}\right]. \label{str}
\end{eqnarray}
The dissipative hydrodynamics of a CFT is determined by only one
kinetic coefficient - the shear viscosity $\eta$. The bulk viscosity
$\zeta$ vanishes for the CFT, while the absence of the particle
number conservation and the use of the Landau frame allow one to
avoid the use of heat conductivity \cite{Jeon:1995zm}. The
dimensional analysis dictates $\eta=F(\lambda)T^d$ where
$F(\lambda)$ is a function of the dimensionless parameters that
characterize the CFT. For strongly coupled CFTs described by an AdS
gravity dual one gets $F=1/\pi$.

\section{The membrane paradigm}

The four laws of black hole thermodynamics are global statements
derived from the Einstein equations restricted to quasi-stationary
perturbations near equilibrium. In the late 1970's and 1980's Damour
\cite{Damour} and later Price, Thorne, and collaborators
\cite{Price:1986yy,membrane} developed a very general analogy
between the local, non-equilibrium physics of any horizon and a
fluid membrane. In this picture the horizon fluid membrane is
governed by the Einstein equations, which correspond to fictitious
Navier-Stokes equations with universal shear and bulk viscosities.
In this section we will first review Damour's approach to the
membrane paradigm, which involves the geometry and dynamics of null
hypersurfaces. Although this formalism is elegant, in some cases it
may be more convenient or conceptually useful to work instead with
Price and Thorne's notion of a \textit{stretched horizon}, a
timelike surface located just outside the true null horizon. We will
briefly discuss how in a particular limit the stretched horizon
becomes the true horizon and the two approaches yield identical
results.

\subsection{Geometry and dynamics of null hypersurfaces}

We will consider a $(d+2)$-dimensional bulk space-time $M$ with
coordinates $X^A, A=0,...,d+1$ with a Lorentzian metric $g_{AB}$.
Let $H$ be a $(d+1)$-dimensional null hypersurface in this
space-time defined by a restriction of the bulk coordinates $F(X^A)
= 0$. The normal vector to this hypersurface, $\ell^A = g^{AB}
\partial_B F$, by definition satisfies
\beq g_{AB} \ell^A \ell^B = g^{AB} \partial_A F \partial_B F = 0
\label{normal} \ . \eeq
This condition implies that for a null hypersurface the normal
vector is also a tangent vector.

We can choose a set of adapted coordinates so that hypersurface in
the bulk space-time is given by $x^{d+1} \equiv r = r_H = const.$,
and denote the remaining coordinates on the horizon surface as
$x^{\mu}, \mu=0,...,d$. In this
coordinate system $\ell^r = 0$, so $\ell^A \rightarrow \ell^\mu$.

The first fundamental form of the horizon is the pullback of the  space-time metric
to the horizon surface.   In the adapted coordinate system,
\beq  \gamma_{\mu \nu} = g_{AB} e^A_\mu e^B_\nu, \eeq
where $e^A_\mu$ represents a horizon basis. It follows from (\ref{normal}) that the horizon metric $\gamma_{\mu \nu}$ is degenerate:
$\gamma_{\mu \nu} \ell^\nu = 0$. We next introduce the auxiliary null
``rigging" vector $m^A$, which is everywhere transverse to the horizon and normalized such that $m^A \ell_A = 1$.  A vector satisfying these conditions is
\beq m^A = m^r = 1, \label{mvector}\eeq
with all other components zero.  Using this vector, one can write  a completeness relation
\beq  \gamma_{AB} = g_{AB} - \ell_A m_B - \ell_B m_A, \label{completeness} \eeq
where on the horizon the $(d+2)$ tensor $\gamma_{AB}$ reduces to the degenerate induced metric.

In order to construct the generalized second fundamental form for a
null surface, consider the space-time covariant derivative
$\nabla_{A}$ projected into the horizon using the transverse
projector $\Pi^A_B = \delta^A_B - m^A \ell_B$
\cite{Gourgoulhon:2005ng} and acting on the normal vector $\ell^A$,
\beq \Pi^C_A \nabla_C \ell^B. \eeq
Since $\ell^B \ell_B= 0$ (\ref{normal}), we have in the adapted
coordinate system
\beq \ell_A \nabla_{\mu} \ell^A = 0 \ . \eeq
This implies that $\nabla_{\mu} \ell^A$ is tangent to the horizon
and can be expanded in the horizon basis
\beq \nabla_{\mu} \ell^A = \Theta_{\mu}{}^{\nu} e^A_{\nu} \ . \eeq
The mixed index object $\Theta_{\mu}{}^{\nu}$ acts as a ``Weingarten map" from the horizon tangent space onto itself and therefore is the
generalized extrinsic curvature of the horizon. Together, these
first and second fundamental forms provide a description of the
embedding of the null hypersurface in the bulk space-time.

Consider Lie transport of $\gamma_{AB}$ along the null normal
vector $\bm \ell$ which is given by the Lie derivative ${\cal
L}_{\bm \ell} \gamma_{AB}$. This expression can be split into
its trace part (the horizon expansion $\theta$)  and trace free part
(the horizon shear $\sigma^{(H)}_{AB}$):
\bea \theta &=& \nabla_A \ell^A \ , \\
\sigma^{(H)}_{A B} &=& \gamma^{C}_{A} \gamma^{D}_B
\nabla_{(C} \ell_{D)} - \theta \gamma_{AB}/d \ \label{shear}. \eea
Here $\gamma_A^B$ is the projector onto the $d$-dimensional spacelike cross-sections of the horizon transverse to $\ell^A$. The components of the Weingarten map are
\bea \gamma^A_D \gamma^C_B \Theta_{C}{}^{D}
&=&
\sigma^{(H)}{}_C^D + \theta \gamma_C^D/d \label{WeinComp1}\\
\Theta_{A}{}^{B} \ell^A  &=& \kappa(x) \ell^B \label{WeinComp2}\\
\Theta_{A}{}^{B} m_B \gamma^A_C &\equiv& \Omega_C \label{WeinComp3}
\eea
Note that in these formulas we have used the covector $m_A$, which is a one-form tangent to the horizon, just as $\ell^A$ is a tangent basis vector. $\kappa(x)$ is the surface gravity and, as we will see below, $\Omega_C$ is a covector whose components can be associated with a horizon ``momentum".  Eqn. (\ref{WeinComp2}) follows just from the null geodesic equation
\begin{equation}
\ell^B \nabla_B \ell^A =
\kappa(x) \ell^A \ ,
\end{equation}
with the surface gravity as the ``non-affinity" coefficient.

We assume that the dynamics of the horizon geometry perturbations
are governed by the Einstein equations, which are (with a non-zero
cosmological constant $\Lambda$)
\beq R_{AB} - (1/2) R g_{AB} + \Lambda g_{AB} = 8\pi T_{AB}^{matt} \
.\eeq
The Ricci tensor contracted with the normal vector and projected
transversely into horizon $R_{AB} \ell^A \Pi^B_C$
can be expressed solely in terms in terms of the intrinsic
horizon metric and extrinsic curvature using a generalization of the
contracted Gauss-Codazzi equation for a null surface \cite{Damour}. In our adapted coordinates
it has the form
\beq R_{AB} \ell^A e^B_\nu = \bar{D}_\mu \Theta_\nu{}^{\mu} - \partial_\nu
\Theta. \label{nullGauss}\eeq
Since the horizon metric is degenerate, one cannot define a unique connection compatible with it. However there is a well-defined rigged covariant derivative operator on the horizon $\bar{D}_\mu$, which is defined in terms of the bulk connection projected transversely into the horizon\cite{Mars:1993mj}.  In the adapted coordinate system it has the form
\beq \bar{D}_{\vec{e}_\mu} e^A_\nu  = \Gamma^{\sigma}{}_{\mu \nu} e^A_\sigma.  \eeq
The right hand side of Eqn. (\ref{nullGauss}) can be expressed as the covariant divergence of a horizon stress tensor
\beq T_{(H)}{}_\nu^\mu = \Theta_\nu{}^{\mu}-\delta_\nu^\mu \Theta. \eeq
From (\ref{WeinComp1})-(\ref{WeinComp3}), we see that the horizon stress tensor has the general form
\beq T_{(H)}{}_\nu^\mu = \kappa m_\nu \ell^\nu + \Omega_\nu \ell^\mu + \sigma^{(H)}{}_\nu^\mu + \frac{1}{d} \gamma_\nu^\mu \theta - \delta_\nu^\mu (\theta+\kappa), \label{genhorstress}\eeq
where we have used $\Theta = \theta+\kappa$.

Consider first the component of (\ref{nullGauss}) along $\bm \ell$, which is the contraction of the Ricci
tensor with $\ell^A \ell^B$. This yields the null geodesic focusing
equation
\beq R_{AB} \ell^A \ell^B = - \ell^\mu \nabla_\mu \theta + \kappa(x)
\theta - \theta^2/d - \sigma^{(H)}_{\mu \nu} \sigma^{\mu \nu}_{(H)}
\ . \label{focusing1} \eeq
Imposing the Einstein equation, there is no contribution from the
Ricci scalar and cosmological constant terms proportional to the
metric due to (\ref{normal}) and we have
\beq  - \ell^\mu \nabla_\mu \theta + \kappa(x) \theta - \theta^2/d -
\sigma^{(H)}_{\mu \nu} \sigma^{\mu \nu}_{(H)} - 8\pi
T^{matt}_{AB}  \ell^A \ell^B = 0. \label{focusing2} \eeq

The other $d$ components of the Gauss-Codazzi equation can be
obtained by projecting (\ref{nullGauss}) orthogonal to $\bm \ell$
using the projection tensor $\gamma^A_B$. When the Einstein equations are imposed, the terms proportional to the metric again do not contribute because by construction $\gamma^A_B$ and $\bm \ell$ are orthogonal. To write this equation Damour splits space and time $(t,x^i)$ by
introducing a horizon basis $\bm \ell =
\partial_t + v^i \partial_i$ and $\bm \partial_i$. The coordinate $t$
parameterizes a slicing of space-time by spatial hypersurfaces and
$x^i$ are coordinates on $d$-dimensional sections of the horizon
with constant $t$. As a result the equations take the form
\beq -{\cal L}_{\bm \ell} \Omega_i - \theta \Omega_i = -\partial_i
\kappa(x) + D_{j} \sigma^{(H)}{}^{j}_{i} + \frac{1-d}{d}
\partial_{i} \theta - 8\pi T^{matt}_{AB} \ell^A e^B_{i}\ , \label{horizonNS}\eeq
where $\Omega_i = \Theta_i{}^{\bm \ell}$. Together the focusing
equation (\ref{focusing2}) and (\ref{horizonNS}) describe the
dynamics of any null hypersurface.

Although the horizon system is an intrinsically relativistic system,
(\ref{horizonNS}) looks just like a $d$-dimensional non-relativistic
Navier-Stokes equation
\beq {\cal L}_{\bm \ell} \mathcal{P}_i + \theta \mathcal{P}_i =
-\partial_i p + 2 \eta D_{j} \sigma^{(H)}{}^{j}_{i} + \xi \partial_{i} \theta - T^{matt}_{AB} \ell^A e^B_{i}, \label{horizonNS1}\eeq
where $\mathcal{P}_i = -\Omega_i/8\pi$ is the membrane's momentum,
$p = \kappa/8\pi$ the fluid pressure, $\eta = 1/16\pi$ the shear
viscosity, $\xi = \frac{1-d}{8\pi d}$ the bulk viscosity, and $\ell^A e^B_{i}
T^{matt}_{AB}$ an external forcing term. Moreover, using the formula
for the expansion as the fractional rate of change in the horizon's
cross-sectional area $A$,
\begin{equation}
 \theta = \cal{L}_{\bm \ell} \ln
\sqrt{\gamma} \ ,
\end{equation}
the focusing equation (\ref{focusing2}) can be
written like a non-equilibrium entropy balance law
\beq \frac{dS}{dt} - \frac{1}{\kappa}\frac{d^2 S}{dt^2} =
\frac{dA}{T} (\xi \theta^2 + 2\eta \sigma^{(H)}_{\mu \nu}
\sigma^{\mu \nu}_{(H)} + T^{matt}_{AB} \ell^A \ell^B), \eeq
where the Bekenstein-Hawking entropy $S = A/4$ and Hawking
temperature $T = \kappa/2\pi$. Therefore there is an complete analogy between the dynamics of a null hypersurface and the dynamics of a non-relativistic fluid.

The viscous entropy production term due to the shear does appear here as one would expect. It is important to note, however, that the general horizon fluid does not actually correspond to a real fluid because it possesses an unphysical negative ``bulk viscosity". Moreover, the second term on the left hand side of the entropy balance law does not appear in hydrodynamics and reflects the general non-local and teleological (as opposed to causal) character of a globally defined null surface.

These discrepancies with hydrodynamics arise because the general membrane paradigm formalism is valid regardless of the size of the Knudsen number $Kn\equiv l_{cor}/L$. For example, the
membrane equations (\ref{focusing2}) and (\ref{horizonNS}) describe
the dynamics of a black hole with spherical topology in an asymptotically flat spacetime (e.g. Schwarzschild). In this case the correlation length of a fluid will scale as $l_{cor} \sim T^{-1}$, where $T$ is the Hawking temperature, while $T^{-1} \sim r_H$, where $r_H$ the horizon radius. Since the horizon is compact, $L$ can be no greater than $\sim r_H$. Thus the dimensionless Knudsen number $Kn$ in these cases is of order unity and hydrodynamics is not an appropriate effective description.

\subsection{The stretched horizon}
\label{stretched}

The above membrane paradigm results can also be obtained via the
stretched horizon formalism (see
\cite{Price:1986yy,membrane,Parikh:1997ma}). Since this approach
employs the familiar formalism of a $(d+1)$-dimensional timelike
surface we can avoid the mathematical complications of dealing with
a hypersurface whose normal vector is also a tangent vector. A
timelike surface is also physically advantageous as a boundary since
a null horizon is an infinite redshift/blueshift surface.

To start, we imagine that the causally complete region of the bulk
space-time $M$ outside the general horizon discussed in the previous
section is foliated by a set of timelike hypersurfaces with
spacelike unit normal vector $n^A n_A = 1$. As the previous section
we can use an adapted coordinate system so that the foliation is
given by surfaces of $x^{d+1} = r = const.$ and the stretched horizon
coordinates are $x^\mu$.  The induced metric on these surfaces is
given by
\beq h_{AB} = g_{AB}- n_A n_B \eeq
and the horizon extrinsic curvature is
\beq K_A^B = h_A^C \nabla_C n^B \label{timeextrinsic}\eeq
One can also consider a unit timelike vector field $U^A U_A = -1$
normal to spacelike $d$ sections of the timelike surfaces. The induced
metric on these horizon cross-sections is
\beq s_{AB} = h_{AB}+U_A U_B. \eeq

The ``distance" between a given timelike hypersurface and the true
null horizon can be parameterized by the affine parameter (or a
function thereof, $\alpha$) along a congruence of ingoing null
geodesics. For example, in the previous section, the set of null
geodesics with tangent vector $m^A$, and affine parameter $r$. The
true horizon is at $\alpha = 0$. To approximate the true
horizon one considers a timelike surface $\alpha \ll 1$ and takes
the limit $\alpha \rightarrow 0$ at the end of calculations. The
combinations $\alpha n^A$ and $\alpha U^A$ are fixed in this limit
such that
\bea \alpha n^A \rightarrow \ell^A \\
\alpha U^A \rightarrow \ell^A.
\eea
In the horizon limit some kinematic quantities will diverge like
inverse powers of $\alpha$ and need to be renormalized so they are
fixed in the limit. For example, one can show that
\bea \alpha K_A^B U^A U_B \rightarrow \kappa(x) \label{limitkappa}\\
\alpha s^C_A s^B_D K_C^D \rightarrow \sigma^{(H)}{}^B_A +
\frac{1}{d} \gamma^B_A \theta \label{limitshearexp} \eea
where $\kappa$ is the horizon surface gravity and $\sigma^{(H)}$ and
$\theta$ are the true horizon shear and expansion
respectively. In contrast,
\beq K_i^A U_A \rightarrow \Omega_i, \eeq
so the horizon's $d$-momentum with respect to $(t,x^i)$  can be
obtained without any $\alpha$ renormalization.

The dynamics of the stretched horizon is determined by the usual
contracted Gauss-Codazzi equations. When the Einstein equation is
imposed we have
\beq D_\nu T^{\mu \nu}_{(S)} = T^{AB}_{\rm matt} e^\mu_A n_B
\label{timelikeGauss}\eeq
where $D_\mu$ is the intrinsic covariant derivative and the
stretched horizon stress tensor is
\beq T^{\mu \nu}_{(S)} = K^{\mu \nu} - h^{\mu \nu} K. \label{horstress}\eeq
Note that this same quantity appears in the
literature as a quasi-local energy-momentum stress tensor. When
evaluated in an asymptotically AdS space-time it is the
Balasubramanian-Kraus stress-tensor \cite{Balasubramanian:1999re}.
After introducing appropriate counterterms to (UV) regularize at the
AdS boundary, $\alpha \rightarrow \infty$, this stress tensor is
equivalent to the stress tensor of the CFT.

In the opposite (IR)
limit, at the horizon, it turns out that (\ref{horstress}) is ill-defined. In this limit the non-degenerate
induced metric $h_{\mu \nu}$ on timelike surfaces becomes the degenerate horizon metric $\gamma_{\mu \nu}$. Therefore the inverse of the horizon metric does not exist and there is no unique canonical way to raise and lower indices. However, one can avoid this problem by always working with the mixed index stress tensor
\beq T_{(S)}{}_\mu^\nu = K_\mu^\nu - \delta_\mu^\nu K. \label{mixedstress} \eeq
The Kronecker delta is well-defined in the horizon limit and Eqns. (\ref{limitkappa}) and (\ref{limitshearexp}) imply that in the
true horizon limit (\ref{mixedstress}) agrees with the horizon stress tensor we defined in (\ref{genhorstress}) from the Weingarten map.
It can also be shown in the same limit that the timelike Gauss-Codazzi equations yield
exactly the null focusing equation (\ref{focusing2}) and the
Damour-Navier-Stokes equation (\ref{horizonNS}).

In the following sections we will apply this general membrane
paradigm formalism to two examples where a large scale hydrodynamic
limit $Kn \ll 1$ exists, black branes in AdS space-time and a
Rindler horizon in flat Minkowski. We will show that in these cases
the focusing and Damour-Navier-Stokes equations are exactly the
relativistic Navier-Stokes equations for a real fluid. Hence the
analogy between null surface dynamics and hydrodynamics is actually
an identity in these cases.

\section{Black branes in asymptotically AdS space-time}

We will first apply the membrane paradigm to a uniformly boosted
black brane in an $(d+2)$-dimensional asymptotically AdS space-time.
This is a solution to the vacuum Einstein equations with negative
cosmological constant
\beq R_{AB} + (d+1) g_{AB} = 0 \ .\eeq
The bulk metric of this unperturbed, equilibrium solution in
Eddington-Finkelstein (EF) coordinates is
\beq ds^2 = -2 u_{\mu} dx^\mu dr + \frac{\pi^4 T^4}{r^2} u_\mu u_\nu
dx^\mu dx^\nu + r^2 \eta_{\mu \nu} dx^\mu dx^\nu,
\label{BBmetric}\eeq
where $T$ is the Hawking temperature and $u^\mu = (\gamma,\gamma
v^i)$ ($\gamma = (1-v^2)^{-1/2}$) is the $(d+1)$-velocity. Note that
the $\mu$ index on the $(d+1)$-velocity is raised and lowered by the
flat metric $\eta_{\mu \nu}$ and its norm with respect to this
metric $u^\mu u_\mu = -1$. The black brane horizon is located at
$r_H = \pi T$ and its normal vector is given by $\ell^r = 0$ and
\beq \ell^\mu = u^\mu. \label{fluidnorm} \eeq

We want to consider perturbations of this black brane horizon
parameterized by allowing $u^\mu(x^\mu)$, $T(x^\mu)$. The horizon
location is $r_H(x^\mu)$ and $x^\mu$ are the coordinates on the
horizon surface. The resulting non-uniform black brane is no longer
a solution to the Einstein equation. However if the velocity and
temperature are slowly varying functions of $x^\mu$ (long
wavelength, long time perturbations) we can solve the Einstein
equations order by order in a derivative expansion and calculate the
corrected metric and stress tensor order by order in Knudsen number.
In what follows we suppose that $u^\mu(\varepsilon x^\mu)$ and
$T(\varepsilon x^\mu)$ and use $\varepsilon$ as a parameter to keep
track of the number of derivatives.

Using (\ref{BBmetric}) we find  the induced metric on the horizon at
zeroth order in $\varepsilon$ is
\beq ds_H^2 = \gamma_{\mu \nu} dx^\mu dx^\nu = (\pi T)^2  (\eta_{\mu
\nu} + u_\mu u_\nu) dx^\mu dx^\nu = (\pi T)^2  P_{\mu \nu} dx^\mu
dx^\nu \label{BBhormet}\eeq
where $P_{\mu \nu}$ is the projection tensor onto the
$d$-dimensional subspace orthogonal to $u^\mu$. Taking the Lie
derivative along $\bm u$ we get the $O(\varepsilon)$ expressions
\bea \theta &=& \partial_\mu u^\mu + d ~ \mathcal{D} \xi
\label{theta1st}\\
\sigma^{(H)}_{\mu \nu} &=&  (\pi T)^2 \left(P^{\alpha}{}_{\mu}
P^{\beta}{}_{\nu}
\partial_{(\alpha} u_{\beta)} - \partial_\gamma u^{\gamma} P_{\mu
\nu}/d \right) \label{shear1st} \ . \eea
We have defined $\mathcal{D} = u^\mu \partial_\mu$ and $\xi = \ln
T$. The horizon shear is equivalent to (up to an overall factor) the
usual fluid shear in hydrodynamics (\ref{str}). Note that (\ref{shear}) implies $\sigma^{(H)}{}^\mu_\nu = \sigma^\mu_\nu$.

We can also use (\ref{BBmetric}) to calculate the other horizon geometrical quantities. From $\Omega_\nu = P^\sigma_\nu m_\mu \nabla_ \sigma \ell^\mu$, we find $\Omega_\mu = -\frac{1}{2} a_\mu$, where $a_\mu = \mathcal{D} u_\mu$.
Using $u^\nu \nabla_\nu u^\mu = \kappa u^\mu$ we find at zeroth order the surface gravity is
\beq \kappa(x) = 2 \pi T(x). \label{surfacegrav0}\eeq

With the kinematical quantities defined, we now consider the
dynamics of horizon perturbations in a derivative expansion in
$\varepsilon$. The dynamics are described by the generalized
Gauss-Codazzi equation (\ref{nullGauss}) in vacuum
\beq \bar{D}_\mu \Theta_\nu{}^{\mu} - \partial_\nu \Theta = 0.
\label{nullGauss2} \eeq
The Weingarten map can be calculated directly from the bulk
covariant derivative along the horizon $\mu$ coordinates
\beq \Theta_\nu{}^\mu = \nabla_\nu u^\mu. \label{Wein}\eeq
We start by calculating the zeroth order (in Knudsen number) part of
this horizon stress tensor. Derivatives of this stress tensor
should yield the equations of ideal hydrodynamics. At this
lowest order we find
\beq \Theta_\nu{}^\mu = -\kappa(x) u_\nu u^\mu. \eeq
Note that this is consistent with the first term (\ref{genhorstress}) since $m_\mu = -u_\mu$. Expanding out (\ref{nullGauss2}) gives
\beq \partial_\nu \Theta_\mu{}^{\nu} + \Gamma^{\beta}{}_{\beta
\lambda} \Theta_\nu{}^\lambda - \Gamma^{\lambda}{}_{\mu \nu}
\Theta_\lambda{}^\mu - \partial_\nu \Theta  = 0,
\label{expandedGauss} \eeq
where, following the definition (\ref{horconnection}), we define the connection as
\beq \Gamma^{\sigma}{}_{\mu \nu} = \frac{1}{2} g^{\sigma \lambda} (\partial_\mu g_{\nu \lambda} + \partial_\nu g_{\mu \lambda} - \partial_\lambda g_{\mu \nu}), \label{horconnection} \eeq
evaluated on the horizon.  Plugging in (\ref{surfacegrav0}) and using the
formula $\Gamma^{\beta}{}_{\beta \lambda} = \partial_\lambda \ln
\sqrt{\gamma}$ we find at $O(\varepsilon)$
\beq - \mathcal{D} T u_\nu - T \partial_\mu u^\mu u_\nu - T a_\nu -
d \mathcal{D} \xi T u_\nu - \partial_\nu T = 0 . \eeq
Projecting along
$u^\nu$ yields
\beq \partial_\mu u^\mu + d \mathcal{D} \xi = 0. \eeq
This equation is equivalent to the vanishing of the horizon
expansion, $\theta = 0$. Note that we could have read off this
equation from the lowest $O(\varepsilon)$ part of the horizon focusing
equation (\ref{focusing2}).

Since we are after the relativistic NS equations we will not
introduce an explicit split of space and time $(t,x^i)$ as done
previously in the membrane paradigm, which would correspond to
hydrodynamic equations written in terms of the $d$-velocity $v^i$
($u^\mu = (\gamma, \gamma v^i)$). Instead, we just project
transverse to $\bm u$ with the operator $P^\nu_\sigma$ to obtain the
other set of $d$ equations. The result is
\beq a_\sigma + P^\nu_\sigma \partial_\nu \xi = 0. \eeq
Together this set of membrane equations is identical to the
equations of relativistic ideal CFT hydrodynamics
\beq \partial_{\nu} T^{\mu \nu} = 0 \eeq
with traceless perfect fluid stress tensor given in (\ref{ideal00}).
The focusing equation is equivalent to the contraction along $\bm
u$, $u_\mu \partial_\nu T^{\mu \nu} = 0$, while the second equation
is equivalent to the projection orthogonal, $P_{\alpha \mu}
\partial_\nu T^{\alpha \nu} = 0$.

An important characteristic of ideal hydrodynamics is that the fluid
entropy current is a conserved quantity $\partial_\mu J^\mu_s = 0$.
In our case the horizon expansion is the fractional rate of change
in the horizon cross-sectional area $\theta =  {\cal L}_{\bm u} \ln
\sqrt{\gamma}$, so our results imply that at $O(\varepsilon)$ the
area is unchanged. Indeed, using the equivalent definition
\beq \theta = \frac{1}{\sqrt{\gamma}}\partial_\mu (\sqrt{\gamma}
u^\mu) = 0, \label{diveqn}\eeq
and the Bekenstein-Hawking entropy formula $S = A/4$, we find
\beq J^\mu_s = \frac{1}{4} \sqrt{\gamma} u^\mu. \eeq
As expected, and in agreement with the entropy current derived in
\cite{Bhattacharyya:2008xc}, the fluid entropy density is
proportional to the horizon area density $(\pi T)^d$.

\subsection{Viscous hydrodynamics}

We now consider $O(\varepsilon)$ terms in the generalized extrinsic curvature and the structure of the membrane equations to $O(\varepsilon^2)$. An immediate difficulty is that a priori the location of the horizon $r_H$ is modified at $O(\varepsilon)$. To compute this correction it seems one would have to know the full $O(\varepsilon)$ metric (the solution to all the Einstein equations up to $O(\varepsilon^2)$\footnote{See, for example, Eqn.
4.24 in \cite{Bhattacharyya:2008jc}}). However, using horizon geometric variables, we can make the following simple argument that the location is unchanged at lowest order. Since the only scalar horizon variable at $O(\varepsilon)$ is the expansion $\theta$, imposing the ideal equation $\theta=0$ means that $r_H = \pi T(x^\mu)+O(\varepsilon^2)$.

Next, using the zeroth order metric (\ref{BBmetric}), and (\ref{Wein}) we obtain
\beq \Theta_\nu{}^\mu = -2\pi T u_\nu u^\mu - \frac{1}{2} a_\nu
u^\mu + \sigma_\nu^\mu + \frac{1}{d} P_\nu^\mu \theta.
\label{1stWein} \eeq
There are several important things to discuss pertaining to this
result.  First, the ideal equations can also be imposed in terms of $O(\varepsilon^2)$ in the membrane equations, so the
last term involving $\theta$ above effectively will not contribute
at this order. Second, the question again arises whether the
$O(\varepsilon)$ corrected metric can affect the remaining three
terms in (\ref{1stWein}). If they do, can we even proceed without knowing the details of these corrections? It turns out $O(\varepsilon)$ corrections in the metric can introduce $O(\varepsilon^2)$ corrections to the shear tensor, but these can only appear at higher order,
$O(\varepsilon^3)$, in the membrane equations. On the other hand,
the corrections to the near-horizon metric will contribute and modify the first two terms of (\ref{1stWein}) at the order we are considering. In particular, from (\ref{genhorstress}) we see that at $O(\varepsilon)$ the non-zero contributions to the horizon stress tensor are
\beq T_{(H)}{}_\nu{}^\mu = - \kappa^{(1)} u_\nu u^\mu + 2\pi T (m^{(1)}_\nu u^\mu - u_\nu \ell^\mu_{(1)}) - \frac{1}{2} a_\nu u^\mu  + \sigma_\nu^\mu + \frac{1}{d} P_\nu^\mu \theta - \delta_\nu^\mu \kappa^{(1)}. \label{genhorstress1st} \eeq
Here $\kappa^{(1)}$, $m^{(1)}_\nu$, and $\ell^\mu_{(1)}$ are the first order parts of the surface gravity, covector, and null normal respectively.

These variables are associated with various ambiguities in the geometrical description that need to be fixed. The first ambiguity arises from the fact that the horizon null normal vector is not unique; any vector obtained by an overall scaling of
the original one is still a null normal vector:
\beq \ell^\mu \rightarrow f(x) \ell^\mu. \eeq
This freedom means that unlike the non-null cases, the generalized horizon extrinsic
curvature is not unambiguously defined (and can be non-symmetric).
Under this scaling \cite{Gourgoulhon:2005ng}
\bea
m_\mu &\rightarrow& f^{-1} m_\mu \\
\kappa' &\rightarrow& f (\kappa + \mathcal{D} \ln f) \label{kappascaling}\\
\sigma' &\rightarrow& f \sigma \\
\theta' &\rightarrow& f \theta. \eea
At zeroth order in $\varepsilon$, $\ell^\mu = u^\mu$ (i.e. $f=1$) is the natural scaling fixed by the equilibrium solution, but at $O(\varepsilon)$  there is again an ambiguity. Since there is a zeroth order $\kappa$ (the temperature), (\ref{kappascaling}) implies that the scaling is equivalent to the freedom to define $\kappa^{(1)}$.
We choose a $f = 1 + O(\varepsilon)$ such that the $O(\varepsilon)$ correction to $\kappa$ is zero; that is
\beq \kappa = 2\pi T + O(\varepsilon^2). \label{kappagauge} \eeq
Thus, fixing this ambiguity determines $\kappa$ at first order and relates the surface gravity to the temperature.

The second ambiguity corresponds to the freedom in the form of the bulk metric evaluated at the horizon, i.e. the choice of $(d+1)$ of the $(d+2)$ functions $Y_B$ in
\beq g_{AB} \rightarrow g_{AB} + \partial_{(A} Y_{B)}. \eeq
One can also see this from the fact that since $m^r=1$ exactly (\ref{mvector}), the covector is $m_\nu = g_{\nu r}$, evaluated at the horizon. We make the gauge choice that
\beq P^\nu_\sigma m^{(1)}_\nu  = \frac{1}{2} (2\pi T)^{-1} a_\sigma. \label{mgauge} \eeq
This sets the $O(\varepsilon)$ correction to $m_\nu$ such that the $a_\nu$ term in (\ref{1stWein}) is eliminated.

The third ambiguity is analogous to the ambiguity in the definition of relativistic viscous hydrodynamics.
This ambiguity corresponds to the
choice in the definition of the fields $T(x)$ and $u^\mu(x)$.
One choice of a hydrodynamic frame is the so called Landau frame,
defined by the $d+1$ conditions imposed on the first order viscous correction to the symmetric fluid stress tensor $T_{(1)}^{\mu \nu}$
\beq u_\mu T_{(1)}^{\mu \nu} = 0 \ . \eeq
These conditions are equivalent to the statement that $u^\mu$ is an eigenvector of the full stress tensor.
Physically the Landau choice is that $T(x)$ and $u^\mu(x)$ are such that in the local rest frame at each point the fluid momentum is zero and energy density can be expressed in terms of equilibrium quantities, without dissipative corrections.

We will fix this ambiguity in the gravitational description by the requirement that
\beq u^\nu T^{(1)}_{(H)}{}_\nu^\mu = 0, \label{Landau2}\eeq
as the analog of the Landau frame choice at the horizon. Imposing this condition, we find the requirement that the first order null normal obey
\beq P^\sigma_\mu  \ell^\mu_{(1)} = 0. \label{ellgauge}\eeq
Note that the frame choice (\ref{Landau2}) implies
the constraint $m_\mu \ell^\mu = 1$ also at first order, i.e.
\beq \ell^\mu_{(1)} u_\mu = m^{(1)}_\mu u^\mu \ , \label{mellconstraint} \eeq
where we use the zeroth order relations $m^{(0)}_\mu=  -u_\mu$ and $\ell^\mu_{(0)} = u^\mu$.

With these gauge and frame conditions (\ref{kappagauge}), (\ref{mgauge}), and (\ref{Landau2}) we can proceed to calculate the
membrane equations using (\ref{expandedGauss}). Imposing the
lower order ideal hydrodynamics equations where they are applicable,
we find
\bea -2\pi T \mathcal{D} \xi u_\nu - 2\pi T \partial_\mu u^\mu u_\nu
- 2\pi T a_\nu + \partial_\mu \sigma^\mu_\nu - 2 \pi T \partial_\nu
\xi - d a_\lambda \sigma^\lambda_\nu - 2 \pi T d \mathcal{D} \xi
u_\nu \nonumber \\ + \Gamma^{(2)}{}^{\beta}{}_{\beta \lambda}
\Theta^{(0)}{}_\nu{}^\lambda - \Gamma^{(2)}{}^{\lambda}{}_{\mu \nu}
\Theta^{(0)}{}_\lambda{}^\mu &=& 0. \label{viscous0} \eea
The last two terms represent corrections to the connection from the
first order metric $\gamma^{(1)}_{\mu \nu}$, which is
composed of first derivatives of  $u_\mu$ and $T(x)$. Since the
zeroth order part of the extrinsic curvature is $2 \pi T u^\mu
u^\nu$, these two terms reduce to
\beq 2\pi T \left(u_\nu \mathcal{D} \ln \sqrt{\gamma^{(1)}} - (1/2)
u^\lambda u^\mu \partial_\nu \gamma^{(1)}_{\mu \lambda}\right). \label{2ndconn}\eeq
The horizon metric must satisfy $\gamma^{(1)}_{\mu \nu} u^\mu = 0$.
Therefore, in terms of the null hypersurface geometric variables it
has the general form (up to an overall factor)
\beq \gamma^{(1)}_{\mu \nu} \sim \sigma^{(H)}_{\mu \nu} + \theta
\gamma^{(0)}_{\mu \nu}. \eeq
In (\ref{2ndconn}) contributions from traceless shear will be zero
identically and in addition we can impose the ideal hydrodynamics
equation $\theta = 0$. Therefore we find both terms vanish.

Contracting the remaining terms in (\ref{viscous0}) with $u^\nu$
yields the scalar equation
\beq \partial_\mu u^\mu + d \mathcal{D} \xi = \frac{1}{2\pi T}
\sigma_{\mu \nu} \sigma^{\mu \nu}.  \label{viscous1}\eeq
One can readily show this equation is just the null focusing
equation up to $O(\varepsilon^2)$. The first order focusing equation
implied $\theta = 0 + O(\varepsilon^2)$. Imposing this result means that
the $u^\mu
\partial_\mu \theta$ and $\theta^2$ terms in (\ref{focusing2}) are
$O(\varepsilon^3)$ and $O(\varepsilon^4)$ respectively. Therefore at
$O(\varepsilon^2)$,
\beq \kappa \theta = \sigma_{\mu \nu} \sigma^{\mu \nu}. \eeq
Using (\ref{diveqn}) this equation can also be expressed as an
entropy balance law
\beq  T \partial_\mu (J^\mu_s) = \frac{\sqrt{\gamma}}{8\pi}
\sigma_{\mu \nu} \sigma^{\mu \nu} = 2 \eta \sigma_{\mu \nu}
\sigma^{\mu \nu} \eeq
where $\eta = \pi^{d-1} T^d/16$ is a shear viscosity. This agrees at
$O(\varepsilon^2)$ with Eqn. B. 27 in \cite{Bhattacharyya:2008xc} and
the shear viscosity is the Kovtun-Son-Starinets \cite{Kovtun:2003wp}
universal viscosity to entropy density ratio $\eta/s = 1/4\pi$ since
$s = (\pi T)^d/4$. Note also that since the right hand side of the
above equation is positive definite, the fluid entropy always
increases. Therefore the second law of thermodynamics for the fluid
is mapped into the area increase theorem of General Relativity.

Projecting with $P^\nu_\sigma$ we find the remaining $d$ equations
are
\beq a_\sigma  +  P^{\nu}{}_{\sigma} \partial_\nu \xi =
\frac{1}{2\pi T} P^{\nu}{}_\sigma (\partial_\alpha
\sigma^{\alpha}_{\nu} - d ~\sigma^{\alpha}_{\nu} a_\alpha).
\label{viscous2}\eeq
The set of equations (\ref{viscous1}) and (\ref{viscous2}) are
equivalent to the projections along and orthogonal to $\bm u$ of the
equations of 1st order viscous CFT hydrodynamics $\partial_\nu
T^{\mu \nu} = 0$, with traceless fluid stress tensor given by
(\ref{visc00}). These results show
that a relativistic CFT fluid flow is encoded in the dynamics of a
black brane horizon in AdS. Specifically, as long as a large scale
hydrodynamic limit exists (so we can expand in derivatives)
\beq \bar{D}_\nu T_{(H)}{}_{\mu}^\nu = \partial_\nu T_\mu^\nu = 0, \eeq
at least up to $O(\varepsilon^2)$.

It is interesting to compare our results with the holographic \textit{fluid-gravity correspondence} in \cite{Bhattacharyya:2008jc}. In this case the black brane solution near the AdS boundary (or equivalently the ($d+1$)-dimensional boundary stress tensor) is expanded in Knudsen number. The Einstein equations at $O(l_{cor}/L)$ projected into this timelike surface are the equations of ideal hydrodynamics and act as constraints on boundary data. In order to obtain the dual $(d+2)$-dimensional solution at $O(l_{cor}/L)$ one can integrate the remaining ``dynamical" Einstein equations into the bulk radial direction, subject to the condition of a regular event horizon. The horizon regularity condition fixes the boundary stress tensor and constraint equations at next order to be those of viscous CFT hydrodynamics, with the particular shear viscosity to entropy density ratio $1/4\pi$. The procedure can be continued to higher orders in $l_{cor}/L$ in the same way and it has already been used to derive the second order transport coefficients. Note, that a
boundary stress tensor with a shear viscosity to entropy density ratio which is different than $1/4\pi$ corresponds to a bulk background
with a naked singularity (see for instance \cite{Gupta:2008th}).

In contrast, our local analysis at the horizon is only applicable at the lowest orders in the expansion in $l_{cor}/L$. At this level, we show that the details of the expanded bulk solution are not required in order to obtain the ideal and viscous hydrodynamics equations. The membrane equations and the implicit condition that the horizon is regular imply a boundary fluid stress tensor with the particular shear viscosity to entropy density ratio
$\eta/s = 1/4\pi$. This is consistent with having a regular bulk.

The horizon dynamics is able to capture the leading long wavelength viscous dynamics of the finite temperature CFT on the boundary,  but clearly
cannot capture the all wavelength dynamics.  In order to go to higher orders in $l_{cor}/L$ using our method would likely require a knowledge of the bulk, possibly via integration out from the horizon into the radial direction.


\section{The Rindler horizon in Minkowski space-time}

To show that our analysis is applicable to cases other than the
black branes in AdS, we consider the example of Rindler space
associated with accelerated observers in $d+2$ Minkowski space-time.
The metric is typically written in the form
\beq ds^2 = -\kappa^2 \xi^2 d\tau^2+ d\xi^2 + \sum^{d}_{i=1} dx'^i
dx'_i \ \label{Rindler1}, \eeq
where $\kappa$ is a constant surface gravity. This metric can be
obtained from the standard Minkowski metric via the coordinate
transformation $x^0 = \xi \sinh(\kappa \tau)$ and $x^{d+1} = \xi
\cosh(\kappa \tau)$. Therefore these Rindler coordinates cover only
a ``wedge" of the full Minkowski space-time. To a uniformly
accelerated observer with worldline $\xi = const.$, the surface
$\xi=0$ is a causal boundary that prevents the observer from an
access to the entire space-time.

To employ the membrane paradigm conveniently, we make a coordinate
change
\bea \xi^2 &=& r/\kappa \\
x'^i &=& \kappa x^i, \eea
where $r$ is an affine parameter along ingoing null geodesics. In
these coordinates the metric (\ref{Rindler1}) has the
Schwarzschild-like form
\beq ds^2 = -\kappa r d\tau^2 + (4 \kappa r)^{-1} dr^2 + \kappa^2
\sum^{d}_{i=1} dx^i dx_i. \eeq
When boosted uniformly in the $x^\mu \equiv (\tau, x^i)$ directions
the metric becomes
\beq ds^2 = -\kappa r u_\mu u_\nu dx^\mu dx^\nu + (4 \kappa r)^{-1}
dr^2 + \kappa^2 P_{\mu \nu} dx^\mu dx^\nu, \label{RindlerSchw}\eeq
with $(d+1)$-velocity $u^\mu$ defined as in the previous section.
Finally, in Eddington-Finkelstein-like coordinates $x^\mu \equiv
(t,x^i)$
\beq ds^2 = -\kappa r u_\mu u_\nu dx^\mu dx^\nu - u_\mu dx^\mu dr +
\kappa^2 P_{\mu \nu} dx^\mu dx^\nu.  \label{RindlerEF}\eeq

The flat Rindler metric is a solution to the vacuum Einstein
equation with zero cosmological constant
\beq R_{AB} = 0. \eeq
To perturb this horizon we allow for, as in the black brane example,
slowly varying (characteristic scale $L \gg \kappa^{-1}$) fluid
velocity $u^\mu(x)$ and $\kappa(x)$. $\kappa$ can be naturally
identified with a temperature in the following way. Unruh
\cite{Unruh:1976db} showed that accelerated observers feel the
quantum vacuum to be a thermal state at temperature $T=a/2\pi$,
where $a$ the observer's proper acceleration. This is essentially a
local temperature and can be expressed in a Tolman form
$T=\kappa/2\pi \chi$, where $\chi= \sqrt{-g_{\tau \tau}} =
\sqrt{\kappa r}$ is the redshift factor. Given this form, we define
$\kappa = 2\pi T(x)$ as the location independent temperature of the
system. The non-uniform Rindler metric is no longer a solution to
the Einstein equation, but as before, one can solve order by order
for the corrections in an expansion in derivatives of $x^\mu$ (small
Knudsen number).

Using (\ref{RindlerEF}) (or (\ref{RindlerSchw})), the zeroth order
induced metric on the horizon ($r=0$) is
\beq \gamma_{\mu \nu} = 4 \pi^2 T^2 P_{\mu \nu}, \eeq
and taking the Lie derivative along $\bm u$ one finds the horizon
shear and expansion have the same forms (\ref{shear1st}) and
(\ref{theta1st}) as in the black brane case. Imposing the membrane
dynamical equations
\beq \bar{D}_\nu T_{(H)}{}_\mu^\nu= 0 \eeq
and following the analysis of the previous section, it is
straightforward to find that these equations are again equivalent to
the equations of relativistic CFT hydrodynamics
\beq \partial_\nu T^{\mu \nu} = 0 \eeq
at least up to $O(\varepsilon^2)$. The shear viscosity $\eta$ is the
same in the black brane case. Therefore assuming the Rindler horizon
has a Bekenstein-Hawking area entropy density, the shear viscosity
to entropy density ratio in the hydrodynamics equations is again
the Kovtun-Son-Starinets ratio $\eta/s = 1/4\pi$.

In this case though, the result cannot be understood as mirroring the
hydrodynamics of a field theory fluid living on an asymptotic
boundary of space-time. For example, the fluid stress tensor cannot
be identified with a boundary CFT stress tensor. Essentially
we have found that this membrane's dynamics can be re-expressed
entirely in fluid mechanical language even without a notion of holography analogous to AdS/CFT. The fluid system here could be interpreted in terms of the near-horizon degrees of freedom as the vacuum thermal state (thermal atmosphere of Rindler particles analogous to Hawking radiation) perceived by accelerated observers \cite{Eling:2008af}.

\section{The non-relativistic limit and incompressible Navier-Stokes equations}

The hydrodynamics of relativistic conformal field theories is
intrinsically relativistic as is the microscopic dynamics. In
particular, the non-relativistic limit of a relativistic conformal
hydrodynamics may not be well defined. Nevertheless, the limit of
non-relativistic  macroscopic motions of a CFT hydrodynamics is
definable and leads to the non-relativistic incompressible NS
equations \cite{Fouxon:2008tb,Bhattacharyya:2008kq}.

On the gravity side we implement this limit by considering the slow
motion regime where the $d$-velocity $v^i$ is a small perturbation.
Temporarily restoring $c$, the horizon coordinates (now split into
space and time) become $(ct, x^i)$ and the non-relativistic slow
motion limit corresponds to $v^i/c \ll 1$. In order to keep track of
the different terms we impose the scaling $\partial_t \sim c^{-2}$,
$v^i \sim \partial_i \sim c^{-1}$ and we consider $c
\rightarrow\infty$. The temperature behaves as $T(x) = T_0(1+ c^{-2}
P(x))$, where $T_0$ is a constant and $P(x)$ is the fluid pressure.
Note this is not the only conceivable slow motion, long distance
scaling limit. For example, one could imagine $v^i$ and $\partial_i$
scaling differently from each other. However, it turns out our
particular scaling is a natural choice because it is a symmetry of
the incompressible NS equations \cite{Bhattacharyya:2008kq}.

Consider the scaling limit of the relativistic
hydrodynamics/membrane dynamics equations derived in the previous
sections, (\ref{viscous1}) and (\ref{viscous2}). At lowest order,
$O(c^{-2})$, the first equation (the focusing equation) reduces
simply to the fluid incompressibility condition
\beq \partial_i v^i = 0. \eeq
The second equation is of $O(c^{-3})$ at lowest order. Under the
scaling the acceleration $a_\sigma \rightarrow \partial_t v_i + v^j
\partial_j v_i$, while the $\xi$ derivative reduces to just a
derivative of the fluid pressure $P$. On the right hand side, the
derivative of the shear tensor contributes $(1/2) \nabla^2 v_i$
after imposing the incompressibility condition. The second shear
times acceleration term does not contribute because it is of higher
order. Therefore we arrive at the non-relativistic NS equation
\beq \partial_t v_i + v^j \partial_j v_i + \partial_i P =  \nu
\nabla^2 v_i, \eeq
with kinematic viscosity $\nu = (4\pi T_0)^{-1}$.

In \cite{Eling:2009pb} we derived the same incompressible NS
equations directly from the membrane focusing equation
(\ref{focusing2}) and Damour-Navier-Stokes equation
(\ref{horizonNS}) discussed above in Section 3 without needing to
know beforehand the fully relativistic hydrodynamics equations. Our
two derivations are completely equivalent. Damour's special horizon
adapted coordinate system and choice of horizon basis ($\ell$, $\bm
e_i = \partial_i$) gives the generic membrane equations their
classic non-relativistic appearance, which earlier allowed us to
easily show they are the incompressible NS equations. The only
difference is that in \cite{Eling:2009pb} we required a surface
gravity
\beq \kappa = 2\pi T_0 (1+P-(1/2)v^2) \eeq
while here we must have $\kappa = 2\pi T_0(1+P)$ in the scaling
limit. The discrepancy can be traced to Damour's parameterization of
fluid velocity $\ell = \partial_t + v^i \partial_i$, while we work
with $\ell = u^\mu = (\gamma, \gamma v^i)$. The null normals differ
by an overall $\gamma = 1 + (1/2)v^2 + \cdots$ factor. As we
discussed above in Section 4, an overall scaling in
the null normal vectors just amounts to a difference in gauge, which
at lowest order just affects how the surface gravity is to be
defined.

\section{Discussion}

In the paper we used the membrane paradigm and applied a Knudsen
number expansion to analyze the $(d+1)$-dimensional horizon dynamics
of a uniformly boosted black brane in a $(d+2)$-dimensional
asymptotically Anti-de-Sitter space-time
and to a Rindler acceleration horizon in $(d+2)$-dimensional  Minkowski space-time. We showed that the horizon
dynamics is governed by the relativistic CFT hydrodynamics
equations. The fluid velocity and temperature correspond to the
normal to the horizon and to the surface gravity, respectively. The
second law of thermodynamics for the fluid is equivalent to the area
increase theorem of General Relativity.

In the Rindler case, unlike the AdS one,
there is no holographic screen
at the asymptotic boundary of space-time.
The result that the
horizon dynamics is governed by the $(d+1)$-dimensional CFT
hydrodynamics equations may imply that the
near-horizon degrees of freedom behave as a conformal fluid.

The derivation of the CFT hydrodynamics equations required a
knowledge of the horizon embedding and employed a local analysis
near this horizon.
The results apply
to a general non-singular causal horizon, as long as there is a
separation between the characteristic scale $L$ of the macroscopic
perturbations and some intrinsic microscopic $l_c$ scale given by
the inverse of the temperature. The non-singularity requirement was
used when contracting the Einstein equations in order to obtain the
membrane equations. The separation of scales was required in order
to have a small Knudsen number and a valid derivative expansion. The
separation of scales does not exist in general. This is the reason
why, for example, in the general Damour-Navier-Stokes equation
(\ref{horizonNS}), the term $\partial_i \theta$ does not vanish and
leads to the assignment of an unphysical negative ``bulk viscosity".

Taking the non-relativistic limit of our analysis, we obtained the
non-relativistic Navier-Stokes equations as found in the membrane
paradigm approach in \cite{Eling:2009pb}. In this way we also
clarified the relation between the surface gravity and the
non-relativistic fluid pressure.

There are various issues to consider next in the membrane paradigm
approach to relativistic CFT hydrodynamics. Most important is whether
the geometrical formulation of the relativistic CFT hydrodynamics as
a hypersurface dynamics can provide a new insight to the nonlinear
fluid dynamics. This is of course relevant also to the
non-relativistic limit, where we obtained the dynamics of
non-relativistic incompressible fluid.

Another important question is the construction of the higher order
derivative terms in the hydrodynamics equations. It is possible
that these terms are still encoded in the membrane equations at
higher orders in $\varepsilon$. However, it is not yet clear if
the analysis can continue to be done locally near the horizon
hypersurface. It is likely a more detailed knowledge of the bulk
space-time will be required \cite{work}.

One can also consider various generalizations of gravity/fluid
correspondence in the membrane paradigm formalism. As a simple
example, since black holes in the presence of an electromagnetic
field act like a charged membrane \cite{Damour,membrane}, it should
be possible to derive the additional hydrodynamic current
conservation equation
\beq \partial_\mu J^\mu = 0 \eeq
for gauge fields in black brane backgrounds \cite{charged}.

Another interesting case to consider is non-conformal hydrodynamics.
The conformal symmetry is broken by non-trivial background matter
fields in the space-time, for example, a scalar field
\cite{Kanitscheider:2009as}. Therefore, one would need to consider
the scalar field equation near the horizon in addition to the
membrane Einstein equations with non-zero matter stresses. The
analysis of the membrane equations should be modified at
$O(\epsilon^2)$, producing a positive bulk viscosity dependent on
the bulk matter field content.

Finally, the membrane paradigm approach may offer an alternative
route to the hydrodynamics equations in generalized gravity
theories, where the Einstein field equations are corrected by higher
curvature terms \cite{Dutta:2008gf} and the shear viscosity to
entropy density ratio of the fluid is no longer simply $1/4\pi$
\cite{viscosity}.

\section*{Acknowledgements}

We would like to thank Y. Neiman for valuable discussions.
The work of Y.O. is supported in part by the Israeli
Science Foundation center of excellence, by the Deutsch-Israelische
Projektkooperation (DIP), by the US-Israel Binational Science
Foundation (BSF), and by the German-Israeli Foundation (GIF). C.E.
is supported by the Lady Davis Foundation at Hebrew University and
by grant 694/04 of the Israel Science Foundation, established by the
Israel Academy of Sciences and Humanities.

\end{document}